\documentclass[conference]{IEEEtran}
\IEEEoverridecommandlockouts
\usepackage{cite}
\usepackage{amsmath,amssymb,amsfonts}
\usepackage{algorithmic}
\usepackage{graphicx}
\usepackage{textcomp}
\usepackage{xcolor}
\def\BibTeX{{\rm B\kern-.05em{\sc i\kern-.025em b}\kern-.08em
    T\kern-.1667em\lower.7ex\hbox{E}\kern-.125emX}}
\begin{document}

\title{A Comprehensive Analysis of Blockchain Applications for Securing Computer Vision Systems}

\author{\IEEEauthorblockN{Ramalingam M\IEEEauthorrefmark{1}, Chemmalar Selvi G\IEEEauthorrefmark{1}, Nancy Victor\IEEEauthorrefmark{1},\\ Rajeswari Chengoden\IEEEauthorrefmark{1}, Sweta Bhattacharya\IEEEauthorrefmark{1},  Praveen Kumar Reddy Maddikunta\IEEEauthorrefmark{1}\\Duehee Lee\IEEEauthorrefmark{2}, Md. Jalil Piran\IEEEauthorrefmark{3}, Neelu Khare\IEEEauthorrefmark{1}, Gokul Yendri\IEEEauthorrefmark{1}, Thippa Reddy Gadekallu\IEEEauthorrefmark{1}\IEEEauthorrefmark{4}\IEEEauthorrefmark{5}\IEEEauthorrefmark{6}}

\IEEEauthorrefmark{1} Vellore Instituite of Technology, Vellore, Tamil Nadu, India.\\

	\IEEEauthorrefmark{2} Department of Electrical Engineering, Konkuk University, Seoul 05029, South Korea.\\
 \IEEEauthorrefmark{3} Department of Computer Science and Electrical Engineering, Sejong University, Seoul, South Korea. \\

  \IEEEauthorrefmark{4} Department of Electrical and Computer Engineering, Lebanese American University, Byblos, Lebanon\\

   \IEEEauthorrefmark{5} Zhongda Group, Haiyan County, Jiaxing 314312, China\\

    \IEEEauthorrefmark{6} College of Information Science and Engineering, Jiaxing University, Jiaxing 314001, China\\

 \IEEEauthorrefmark{6} Division of Research and Development, Lovely Professional University, Phagwara 144401, India\\

 }

\maketitle

\begin{abstract}
Blockchain (BC) and Computer Vision (CV) are the two emerging fields with the potential to transform various sectors.The ability of BC can help in offering decentralized and secure data storage, while CV allows machines to learn and understand visual data. This integration of the two technologies holds massive promise for developing innovative applications that can provide solutions to the challenges in various sectors such as supply chain management, healthcare, smart cities, and defense. This review explores a comprehensive analysis of the integration of BC and CV by examining their combination and potential applications.  It also provides a detailed analysis of the fundamental concepts of both technologies, highlighting their strengths and limitations. This paper also explores current research efforts that make use of the benefits offered by this combination. The effort includes how BC can be used as an added layer of security in CV systems and also ensure data integrity, enabling decentralized image and video analytics using BC. The challenges and open issues associated with this integration are also identified, and appropriate potential future directions are also proposed. 

\end{abstract}


\section{Introduction}  
Computer vision (CV) is rapidly expanding and gaining significant prominence in the field of Artificial Intelligence (AI) \cite{su2021affective},\cite{fisher1994computer}. It involves the development of algorithms and techniques that enable computers to interpret and understand the visual world, including images and videos \cite{haralick1991glossary}. CV replicates the human visual system by utilizing sensing and interpretation devices that function similarly to the human eyes and visual cortex in the brain \cite{jarvis1983perspective},\cite{szegedy2016rethinking}. The significance of CV in today's world is increasing with the availability of data and high computing power. Thus, CV has the potential to revolutionize the way we interact with machines and the world around us \cite{zhang2023computer},\cite{voulodimos2018deep}. 

Today, CV has numerous applications across a wide range of industries, including healthcare \cite{bentadj2023computer}, security \cite{sage1999security}, entertainment \cite{hou2023aesthetics}, defense \cite{janakiramaiah2023military}, self-driving vehicles \cite{de2023predicting}, disaster relief and emergency \cite{daniel2023wildfire}, modern agriculture \cite{gupta2023artificial}, banking industry \cite{mohapatra2023implementing}, manufacturing industry \cite{wang2023knowledge} and robotics \cite{maroto2023active}. CV has evolved significantly over the years, with different techniques and approaches being developed to address the challenges of interpreting and understanding visual data \cite{huang1996computer}. In the 1950s and 1960s, researchers began developing techniques for image processing and pattern recognition, including edge detection, thresholding, and template matching \cite{ruaro2005toward}. These early techniques laid the foundation for later work in CV. In the 1970s and 1980s, researchers began developing feature-based methods in CV including corner detection, edge detection, and texture analysis \cite{li2015efficient}. These methods focused on identifying and extracting features from images that could be used to identify objects and patterns.     

In 1980s and 1990s, researchers began developing geometric methods for CV, including stereo vision, optical flow, and shape from shading \cite{blicher1985edge}. These methods focused on using geometric principles to infer the 3D structure of objects from 2D images. In 2000s and beyond, the rise of machine learning and deep learning revolutionized the evolution of CV \cite{vodrahalli20173d}. These techniques enabled researchers to develop algorithms that could learn from visual data and automatically extract features from images, enabling more sophisticated and accurate applications. Overall, CV has become a critical tool in today's research across a wide range of fields. As CV continues to evolve, it is likely to play an increasingly important role in research, driving innovation and discovery in various applications.
\begin{table}[t]
\caption{List of Key Acronyms.}
\centering
\begin{tabular}{|l|l|}
\hline
\textbf{Acronyms} & \textbf{Description}              \\ \hline
CV                & Computer Vision                   \\ \hline
BC                & Blockchain                        \\ \hline
AI                & Artificial Intelligence           \\ \hline
MIA               & Medical Image Analysis            \\ \hline
5G                & Fifth Generation                  \\ \hline
ML                & Machine Learning                  \\ \hline
CNN               & Convolutional Neural Network      \\ \hline
RNN               & Recurrent Neural Network          \\ \hline
API               & Application Programming Interface \\ \hline
UAV               & Unmanned Aerial Vehicle           \\ \hline
IoT               & Internet of Things                \\ \hline
CCTV              & Closed Circuit Television         \\ \hline
\end{tabular}
\end{table}

\begin{table*}[t]
\centering
\caption{Comparison highlights between previous surveys and the current work}
\label{tab:my-table}
\begin{tabular}{|p{1 cm}|p{2 cm}|p{6cm}|p{5cm}|}

\hline
Research & Application Domain &  Contributions &  Limitations \\ \hline
 \cite{ tian2020computer} & Agriculture &
  The advancement of CV in smart agriculture including crop growth monitoring, crop disease prevention using pesticides, automatic crop harvesting, quality inspection of crop products. &
  
  The key challenges faced by the use of CV technology in automating smart agricultural solutions.
  More emphasis is given to the learning model in automating intelligent agriculture,  however, how CV can contribute its fullest potential toward smart agricultural decision-making process is missing. \\ \hline
 \cite{ kakani2020critical} &
 Food Industry &
  The application of CV technologies in the food industry by discussing state-of-the-art methods. Different usecases related to agricultural start-ups using AI and CV technologies. &
  Although the recent trends of applying AI and CV technologies have been discussed well, the viability of developing it in a real-time setup has not been justified. \\ \hline
 \cite{ esteva2021deep} &
 Healthcare &
  Modern CV technologies for enhancing medical applications.  Open problems for deploying  such real-time clinical solutions with key challenges and  research directions. &
  Privacy and security aspects of medical patient data was not considered when explaining the role of CV in medical applications. \\ \hline
\cite{ cheng2021fashion} & Fashion &
  Intelligent fashion techniques using the unprecedented growth in CV applications. The open problems and research opportunities in fashion recommendation. &
  Key challenges in developing the intelligent complex fashion solutions rely on  huge computational resources are not explored. \\ \hline
 \cite{ gonzalez2023survey} & Underwater applications &
  Underwater CV applications with its use case and research challenges. &
  Emphasizes only on certain image processing techniques to enhance underwater object detection models which is too specific. \\ \hline
NA & Our Paper &
  The first-of-its-kind survey on applications of BC in CV.  It highlights the role of BC in CV with possible applications . Presents the major challenges and future research scope of applying BC in CV. &
  NA \\ \hline
\end{tabular}%

\end{table*}

\subsection{Highlights of the review}

In this subsection, the background study of existing survey papers related to CV are discussed. Tian et al \cite{tian2020computer} provided the comprehensive overview of the current state of CV technology in agricultural automation, its potential benefits, and the challenges and limitations that need to be addressed for wider adoption of CV in the agricultural sector. They introduced the concepts of agricultural automation and its potential benefits such as increased efficiency, reduced labor costs, and improved crop yields. The importance of role of CV technology in agricultural automation was discussed and reviewed various applications of CV technology in agricultural automation, including crop monitoring and analysis, fruit and vegetable grading, livestock monitoring, and weed detection. Finally, they concluded with the challenges and limitations highlighting the need for more research in this area, particularly in the development employing machine learning algorithms for analyzing agricultural data.

Kakani el al \cite{kakani2020critical} presented an exhaustive survey of how CV and AI are being used in the food industry, cover a range of areas such as quality control, food safety, food processing, and packaging. They analyzed the latest advancements in CV and AI techniques, such as deep learning, machine learning, and image processing, and assess their effectiveness in tackling the obstacles that the food industry is confronting. Finally, they concluded that CV and AI have great potential to improve the efficiency, quality, and safety of the food industry and also highlights the need for huge demand of research exploration in this area.

Estava et al \cite{esteva2021deep} provides an overview of recent advances in deep learning techniques applied to medical CV. They covered a broad range of topics, including medical image segmentation, registration, classification, and detection, as well as other applications such as medical robotics and pathology. They discussed about the key challenges and opportunities presented by deep learning methods in medical CV, as well as the potential for improving clinical outcomes and patient care through these techniques. Finally, they concluded the paper by highlighting the constraints and areas for improvement, along with the future research directions concerning deep learning approaches in medical CV.

Cheng et al \cite{cheng2021fashion} presents an extensive review of the latest developments in CV techniques as applied in the fashion industry. They explained the various areas of CV in fashion industry such as fashion image retrieval, segmentation, classification, detection, and generative modeling. They explored the potential advantages and challenges posed by CV techniques in the fashion industry, as well as the prospects of enhancing consumer retail and e-commerce experiences. They concluded the paper by emphasizing the limitations and areas for future research to be explored for the potential role of CV towards the growth of fashion industry.

Gonzala et al \cite{gonzalez2023survey} presents a comprehensive overview of the latest advancement in CV techniques as applied in underwater environments. They presented a broad range of areas applying CV in underwater settings, including image enhancement, restoration, segmentation, tracking, and object recognition. They explored the potential challenges and benefits using CV techniques in the underwater domain, and also advancing the scientific research, marine conservation, and industrial applications. They finally concluded the paper by highlighting the limitations and future research scope for the role of CV in underwater research.

Thus, this subsection aimed at comparing the proposed survey work against the existing survey works which are summarized in Table \ref{tab:my-table}. Each of this survey articles has discussed about the recent advancements in the CV technologies, state-of-the-art techniques, domain applications, case studies, open challenges and future directions related to CV technology in different application domains. The insights gained from these survey articles motivated us to write this extensive survey paper highlighting how CV can expand its evolution using BC,  the need for integrating BC in CV like data security, data sharing and distributed training, motivation of BC for CV with different CV enabling technologies and ultimately discussing the key challenges to be addressed with the future research directions. Hence, this was the main motivation to initiate a first-of-its-kind study focusing on BC applications in CV models. The main contributions of this survey are summarized as follows:
\begin{itemize}
    \item A first-of-its-kind survey is presented on BC-based platforms for deploying CV applications.

    \item Discussion on various CV enabling technologies are presented.
    
    \item An exhaustive review on the applications of BC using CV technologies were discussed.

  \item Finally, discussion of research challenges and future perspectives related to the integration of BC for CV-based applications are highlighted.
\end{itemize}

\begin{figure*}[t]
	\centering
	\includegraphics[width=.6\linewidth]{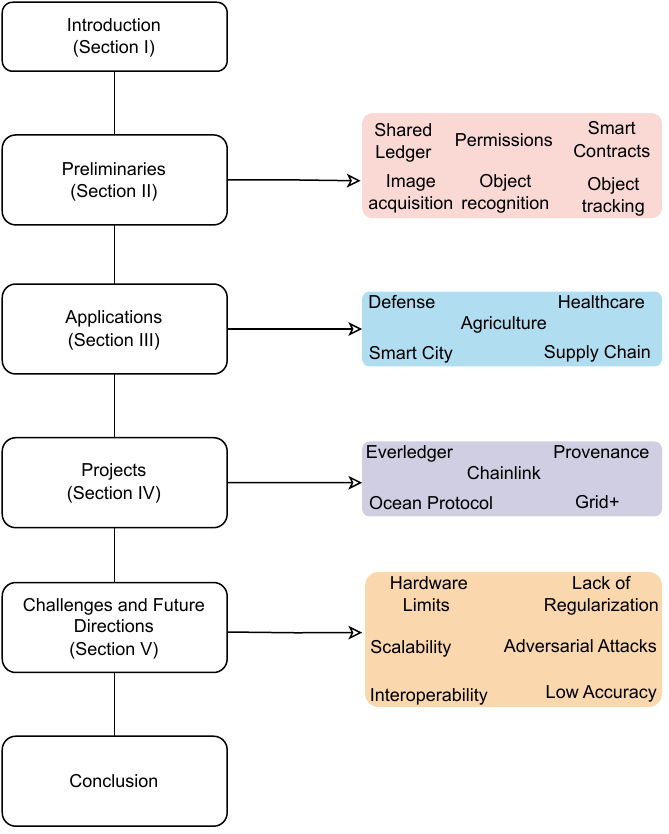}
	\caption{Schematic outline of the paper}
	\label{Outline}
\end{figure*}

The rest of the paper is structured as follows. Section II presents the CV introduction and its significant role, CV enabling technologies and motivation of integrating BC with CV. Section III discusses the critical applications of BC for CV, and section IV highlights the major challenges and future research scope. Finally, the paper is concluded with the open research problems and benefits of integration of BC for CV. Fig.~\ref{Outline} depicts the schematic outline of this study.

\section{Preliminaries}

The following subsection will discuss in detail about the working process of CV and BC. 
\subsection{BC}
BC technology enables the decentralized and distributed information sharing system for transmitting the immutable data in an encrypted and secured way. It acts as an immutable ledger that enables recording of transactions and asset tracking in a business network. Any asset of value can possibly be traded and tracked in a blockchain network thereby reducing risks and cost of resources. BC is basically a technology that is based on bitcoin cryptocurrency which is a public ledger systems that ensures integrity of the transaction data \cite{peters2015trends,al2019blockchain,karame2012two}. Although bitcoin is the most predominantly used application of BC but it has the potential to be applied in versatile applications beyond the scope of cryptocurrencies. Thus, finance, healthcare, manufacturing, distribution and government administrations are just a few of the industries where BC has permeated a wide range of applications. A BC application consists of small unit of tasks that are stored in public records. The blocks get executed, implemented and stored in the BC to achieve validation from all miners who are part of the BC network allowing each transactions to be reviewed but not updated thereby facilitating transactions in a decentralized fashion. The major characteristics of BC include decentralization, persistency and anonymity \cite{casino2019systematic,karame2012two,bhattacharya2022blockchain}. 

The name "blockchain" refers to the way transaction data is stored—in blocks that are connected in a chain. The BC grows with the increase in the number of blocks wherein the timings and sequence of the transactions get recorded and confirmed. Finally, these are entered into a distinct network that is governed by rules accepted by network members. The blocks in a BC consists of three major components - header, data section and finally the hash. The header includes meta data, including the hash of the preceding block and a timestamp with a random integer to be used in the mining process. The data section  includes information pertaining to transactions and smart contracts  which gets stored in the blocks. The blocks in the BC contains a digital fingerprint  or an unique identifier termed as hash. It also keeps the hash of the preceding block and the timestamped batches of recent valid transactions. The prior block hash connects the blocks and stops any block from being changed or from being introduced in between two already existing blocks which renders a tamper-proof BC \cite{zheng2017overview}.

\begin{figure*}[t]
	\centering
	\includegraphics[width=.85\linewidth]{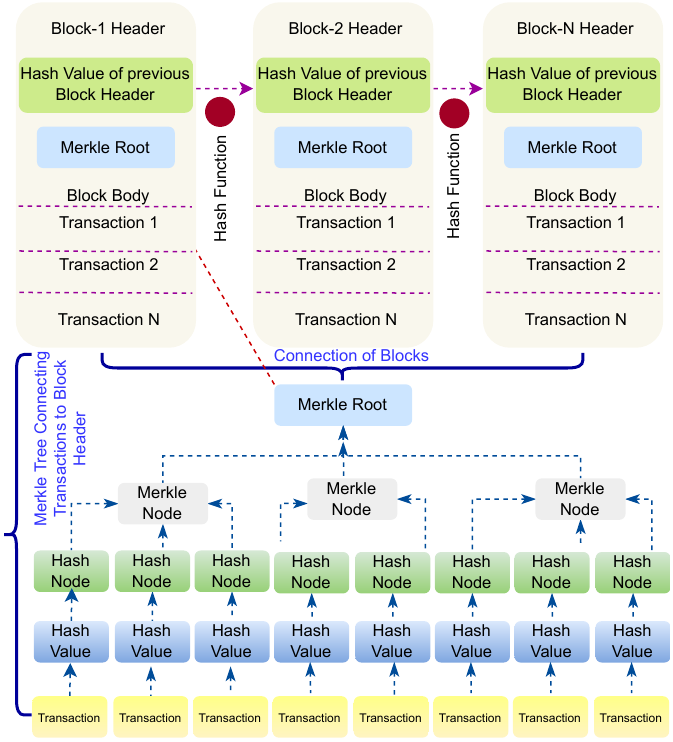}
	\caption{Working Process of BC}
	\label{Working_Process_of_BC}
\end{figure*}

The functioning of the BC framework relies on four key aspects namely the shared ledger, permissions, smart contracts and the consensus. A distributed system of records shared over a business network is known as a shared ledger and allows for "append-only" updates. The shared ledger allows transactions to be recorded only once eliminating chances of duplication of efforts which are atypical of any traditional business networks. Permissions make sure that transactions are protected, verified, and authorised. With the capacity to limit network membership, organisations can more easily abide by data protection laws like those outlined in the EU General Data Protection Regulation (GDPR) and the Health Insurance Portability and Accountability Act (HIPAA). Smart contract is a set of rules or agreement that controls business transactions. It is automatically carried out as part of a transaction and is stored on the BC \cite{deepa2022survey,mohanta2018overview}. The consensus enables all stakeholders agree to the network-verified transactions. Various consensus mechanisms are used as part of BC frameworks namely proof-of-stake, multi-signature and Practical Byzantine Fault Tolerance (PBFT)\cite{hakak2021recent}. The simplified architecture of BC and its components is illustrated in Fig.~\ref{Working_Process_of_BC}.

BC and CV are two major innovations in this "information age" of technology advancement wherein the combined potential of both of these technologies are enormous. Visual data can be captured using CV systems, in the form of pictures or video which can be further recorded and verified using BC technology. This ensures security of the data enabling confident sharing and accessing of the same by authorized stakeholders. The combination of BC and CV can thus help in addressing diversified issues pertaining to data ownership and privacy across versatile domains. 

\subsection{CV}

CV is the study of enabling computers to interpret and understand visual information from the world around us. In order to achieve this goal, a typical CV system consists of multiple components and stages that work together to process and interpret visual data \cite{orhei2020end,chowdhary2021computer}. The anatomy of CV can be broken down into key stages such as image acquisition \cite{sturm2011camera}, preprocessing \cite{wasza2011real}, feature extraction \cite{papakostas2005efficient}, object recognition \cite{de2019does}, tracking \cite{wu2013online}, and interpretation \cite{deguchi2000direct}. Each of these stages plays a critical role in enabling computers to make sense of visual data and extract useful information from it. Understanding the anatomy of CV is essential for researchers and practitioners working in this field, as it provides a framework for building and optimizing CV systems for a wide range of applications. 

Following are some essential components of a standard CV system: 1. Image acquisition: The first step in any CV system is to acquire images or video data using cameras or other imaging devices \cite{north2006seeing}. 2. Preprocessing: Once the image data has been acquired, it may undergo preprocessing to enhance or filter the data. For example, noise reduction or image normalization can be applied in this process \cite{solsona2017new}. 3. Feature extraction: In this stage, the CV system extracts features from the image data that are relevant to the task at hand. This might include edges, corners, or other visual patterns \cite{azhar2015batik} . 4. Object recognition: Based on the extracted features, the system attempts to recognize objects in the image data. This may involve using machine learning algorithms to compare the extracted features to a database of known objects \cite{zou2023object}. 5. Object tracking: Once objects have been identified, the system may track their movement over time, either within a single image or across multiple frames of video \cite{guo2021learning}. 6. Interpretation: Finally, the system interprets the results of the previous stages and produces an output based on the task at hand. This could be anything from labeling objects in an image to guiding a robotic system \cite{marvcivs2023effect}.

\begin{figure*}[t]
	\centering
	\includegraphics[width=\linewidth]{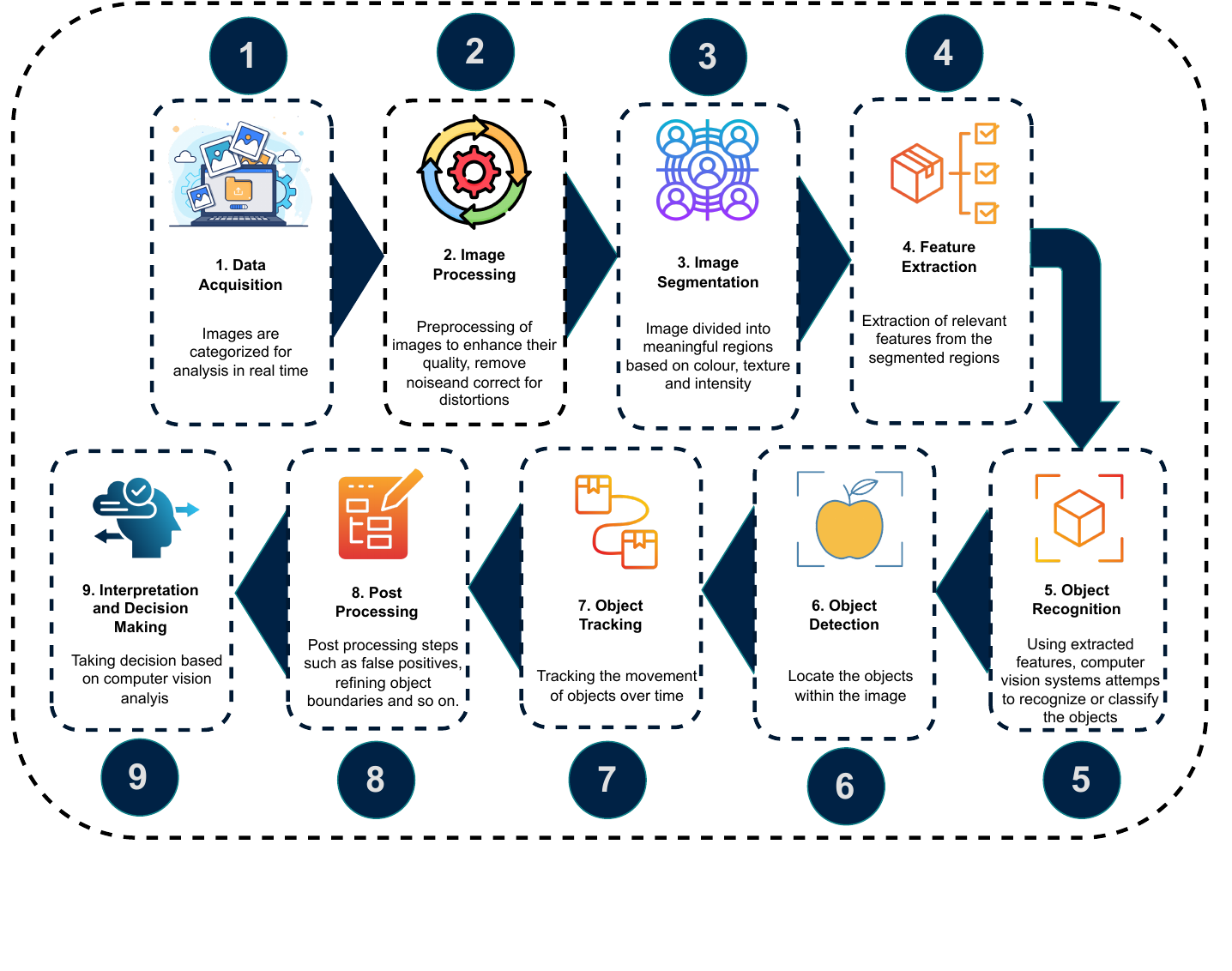}
	\caption{CV working process}
	\label{Working Process of CV}
\end{figure*}

\begin{figure*}[t]
	\centering
	\includegraphics[width=\linewidth]{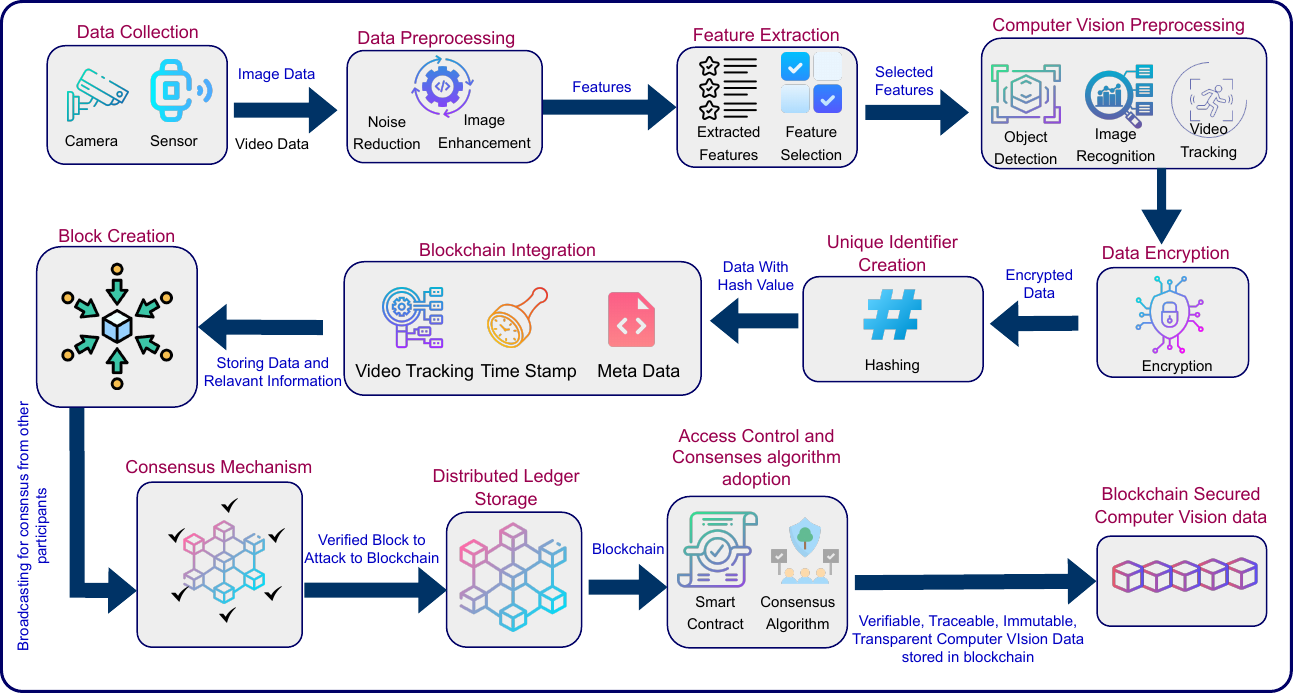}
	\caption{Integration of BC with CV}
	\label{BC_and_CV_Integration}
\end{figure*}
CV focuses on making it possible for machines to perceive and comprehend the visual data from the outside environment. The representative features of CV are image acquisition, image processing, feature extraction, object recognition, and scene interpretation. Image acquisition is the first step in the pipeline of building models based on CV. This is the process of acquiring images from the hardware-based sources such as digital cameras and infrared sensors. The quality of acquired images rely on numerous factors like lighting conditions, camera resolutions and sensor configurations \cite{schindewolf1994evaluation}. Hence, image processing is the pivotal step in CV to preprocess and extract the key features from the acquired images. Filtering \cite{jampani2016learning}, Image Segmentation \cite{singh2010study} ,Feature extraction \cite{bhargava2021fruits}, image registration \cite{brown1992survey} and image restoration \cite{zoran2011learning} are typical Image Processing techniques. The working process of CV is illustrated in Fig.~\ref{Working Process of CV}.

In summary, the structure of CV offers a complete understanding of building and refining systems capable of interpreting and comprehending visual data. Hence, it is considered as a crucial area of research and innovation with immense potential for a diverse array of applications.

\subsection{Motivation}

BC technology has the potential to bring significant changes to various industries. By integrating BC in CV, there are enormous opportunities to enhance potential of CV in several ways: 

1. Data Security: CV handles huge amount of sensitive data which are mostly used in critical applications such as medical, defense, and banking to train the black box models. When BC is integrated with CV, BC can create a transparent and tamper-proof system that stores and verifies all data generated by CV systems. This means that any tampering or unauthorized access to data can be immediately identified and traced back to its source. Thus, such extremely sensitive data produced from CV systems when stored using BC, enables more secured transaction of sensitive data over the distributed network of nodes promoting hacker-resistant data transmission.  

2. Data Sharing: CV involves huge amount of data to be accessed in order to develop robust learning model.  By using a BC-based identity verification system, users can be sure that only authorized individuals are accessing and using data. For example, a smart contract could be used to automatically grant access to authorized users who meet certain criteria and helps to mitigate the risks of data breaches, identity theft, and other security threats. Hence, the use of smart contracts and identity verification can further strengthen the security of data sharing, ensuring that only authorized individuals can access and use the data generated by CV systems.

3. Distributed Training : CV develops models based on deep learning algorithms which require heavy computational resources to train the models. When such models are deployed based on the BC-configured platforms, high computational resources can be distributed by multiple parties for the training process to perform cost-effective and high accuracy robust models. Once the CV models are trained, they can be stored on the BC network, making them easily accessible to all the parties involved in the training process. Therefore, the integration of BC technology can enable distributed training of CV models, leading to significant improvements in the efficiency and scalability of the training process. Fig.~\ref{BC_and_CV_Integration} has given the illustrated view of how BC can be integrated into CV process. 

To summarize, integrating BC technology with CV offers significant potential in transforming the way data is managed and protected. By establishing a decentralized, transparent, and tamper-proof data management system, BC can enhance data security and privacy while also improving the accuracy and reliability of CV systems. Smart contracts and identity verification can be leveraged to further bolster data security, ensuring that only authorized users have access to the data generated by CV systems. As CV continues to gain widespread adoption across multiple industries, integrating BC technology can facilitate increased trust, transparency, and innovation within the data ecosystem. 

\begin{table*}[t]
\centering
\caption{Feature comparison table of standalone CV systems with BC secured CV systems }
\label{tab:Features}
\begin{tabular}{ |p{0.5cm}|p{1.2cm}|p{7cm}|p{7cm}|}
\hline
\textbf{Sl.No} & \textbf{Feature}  & \textbf{BC secured CV systems}                                                    & \textbf{Standalone CV systems}                             \\ \hline
1.             & Security          & Improved through immutability, cryptography and decentralized consensus approach. & Relies on Traditional security measures                    \\ \hline
2.             & Data Transparency & Through public ledger, it provides auditable access to visual data                & Depends on implementation and architecture                 \\ \hline
3.             & Data Integrity    & Immutable records of image data, preventing authorized manipulation               & Integrity measures may vary based on system implementation \\ \hline
4.             & Decentralization  & Distributed network make sure no central authority controls the data              & Centralized control and authority of data may be possible  \\ \hline
5.             & Trust             & Through consensus approach and cryptographic protocols                            & May rely on reputation and reliability of the system       \\ \hline
6.             & Privacy           & BC offers privacy through encryption and selective data sharing                   & Privacy depends on implementation and configuration        \\ \hline
\end{tabular}
\end{table*}

\section{BC and CV integration enabled Applications}

\begin{figure*}[t]
	\centering
	\includegraphics[width=\linewidth]{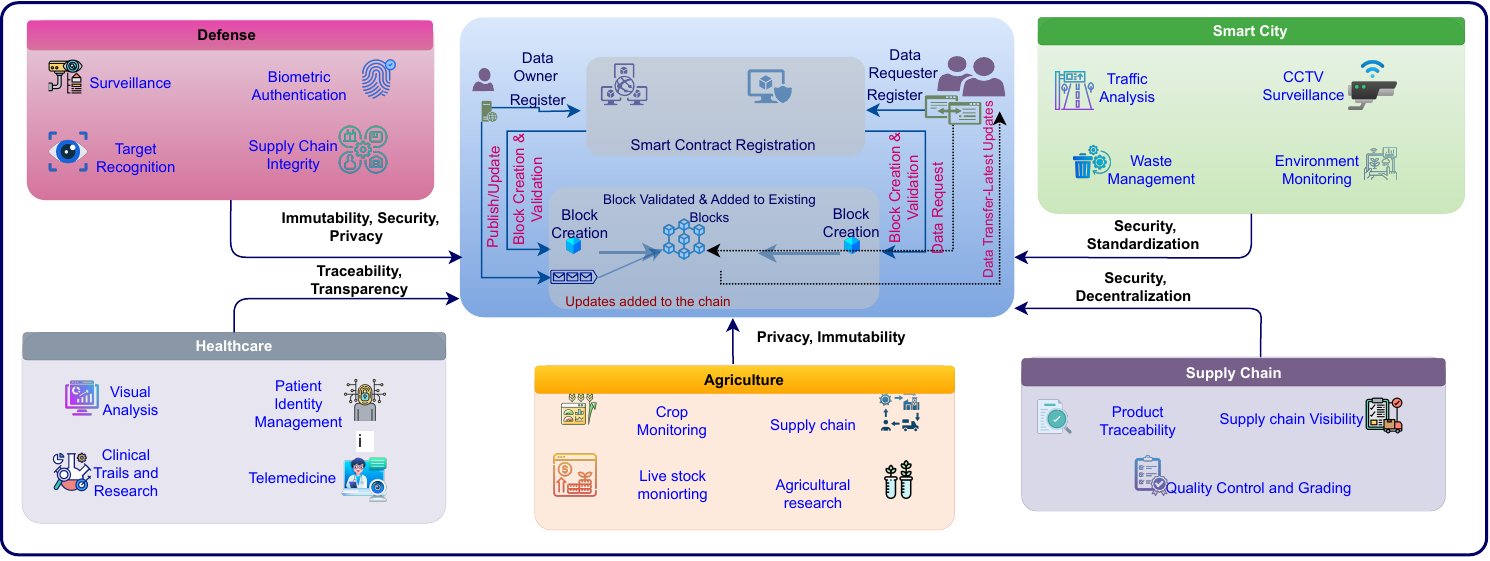}
	\caption{BC and Computer Vision Enabled Applications.}
	\label{fig:BC and CV Enabled Applications}
\end{figure*}

The field of CV has rapidly grown and transformed various industries, but ensuring the authenticity and reliability of the data used to train and test algorithms remains a persistent challenge. One promising solution to this challenge is the use of BC technology, which can provide a secure and transparent framework for managing data in CV applications. With BC, CV algorithms can be trained and tested on tamper-proof data, ensuring the accuracy and robustness of the system. Additionally, BC can enable secure and efficient data sharing and collaboration among stakeholders in the CV ecosystem. This section aims to explore the potential of BC technology in CV applications, highlighting its benefits, challenges, and possible use cases. Fig.~\ref{fig:BC and CV Enabled Applications} has given the illustrated view of the applications
that will benefit from integrating BC and CV and Table \ref{tab:Features} has given the comparison analysis based on the features of traditional CV systems with BC secured CV systems.

\subsection{Defense sector}

In today's world, security has arisen as a significant concern. The defense sector has benefited from CV systems in many ways, including autonomous vehicles, tracking, target recognition, and surveillance. CV systems are predominantly utilised as unmanned aerial vehicles (UAVs) for surveillance in numerous military operations. Though soldiers monitor vulnerable areas manually, it is essential to integrate and deploy advanced technologies, such as drones and surveillance cameras \cite{ding2018amateur}, to surveil areas where continuous monitoring may not be feasible. The authors in \cite{sharma2020wearable} have discussed the significance of technologies in today's defense sector. With the support of technology, instances can be recorded and used as evidence for any future investigation. UAVs, commonly known as drones, are broadly used in several industries, including military services \cite{hadi2023comprehensive}. They are particularly beneficial for monitoring regions that are not reachable by humans. Although conventional drones are not capable of making decisions based on incidents, intellectual drones, or UAVs \cite{matthew2021artificial} can take action based on happenings in the surveillance area and transmit information immediately to the control room.

Implementing CV systems carries various challenges for the defense sector, including integration with existing systems, quality and quantity of data, cost, security, and adaptability. Among these challenges, security is the primary issue that the defense sector should address. This sector demands a high level of security, and CV systems deployed for monitoring should be secured to prevent unauthorized access or manipulation of data. Additionally, CV systems may be vulnerable to cyber-attacks, which could compromise sensitive information of the defense sector.

Intellectual cameras, which make use of CV techniques for visual data analysis and management, have enormous applications in the defense sector. These applications include border security, threat detection, facial recognition, access control, drones, and unmanned aerial vehicles. However, despite these applications, a crucial issue remains unresolved. CV systems must ensure data privacy, secured data storage, data authenticity, and improved traceability while storing the data. These requirements while storing CV data can be resolved by the use of BC in CV systems, providing an added layer of security to the analyzed data.

Drone technology is indisputably beneficial for defense, but it is also susceptible to hacking \cite{yaacoub2020security} \cite{westerlund2019drone}. A hacked drone does not essentially need to be physically captured; hackers can hack the camera of the drone and steal any sensitive information captured from a battle zone, war zone, or any critical area under monitoring. The main issue with drone technology is that it can be remotely operated and its actions depend on wireless communications. Though programming languages used for developing drone software were originally generated to manage and control drones, there were flaws in the languages used that make it easy for hackers to hack the software and take control of the drone. These susceptibilities can result in the loss of information and human lives also, which can be a serious threat to the nation. So, during surveillance through cameras, the primary issue that needs to be addressed is the authentication and security of drones \cite{rana2019intelligent} during their flight \cite{cheon2018toward} \cite{srinivas2019tcalas}.

To secure a drone or UAV, an intellectual approach is required to achieve two goals: 1) prevent unauthorized access to the device and 2) safeguard the information stored in the device. In connection with this, the authors have proposed an intelligent approach for securing the drone or UAV using BC technology. The authors in \cite{9573572} have used image collecting and sensing by drones and UAVs with BC security to create a secured mechanism by encoding the files in the device using hash technology. Timestamp information along with GPS information is used to keep a record of transactions between the server and the drone. The captured information is converted into a hash value using the hash function, and the hash values were encrypted using a private key to create a digital signature. The receiver has used the same hash function on the received information to generate the hash value, and the value generated at the receiver's side should match the hash value received from the sender. This mechanism helps ensure the validity of the digital signature indicating the genuineness of data inside the drone. The proposed mechanism has been tested on consumer-friendly drones, with the server operating on mobile. This mechanism offers reliable security for the data and prevents unauthorized access to the device.

In military surveillance operations, drones are playing a crucial role, and their use is expected to continue to rise in today’s society. However, from a technical perspective, though the use of drones has many benefits, the design and deployment of drone technology present its own set of challenges \cite{kumar2019analysis}. These challenges arise due to the variety of topologies drones can operate in, unstable connections \cite{kumar2019analysis} \cite{namuduri2012airborne}, and most importantly, inadequate security \cite{yazdinejad2020enabling}. To address these security issues, the authors in \cite{9163144} have proposed an architecture that divides the shadowing area into multiple zones, with each zone to a drone controller responsible for activities such as authentication, movement between zones, and communication with other drones using a distributed ledger powered by BC technology. The proposed method allows for the migration of drones between zones while ensuring that the data recorded in each zone is secure. The authors verified the proposed method in a smart city environment by dividing the city into multiple zones, and it helped to confirm low latency and secure communication during implementation.

\subsection{Healthcare sector}

The new generation of information technology has transformed healthcare into smart healthcare. Smart healthcare is a multifaceted, all-encompassing improvements rather than merely a basic technological advancement \cite{chen2023information}. Now the medical model shifted their focus from disease-oriented to patient-oriented care. Similarly, the construction of information technology has shifted from clinical information to regional medical information, medical management changed from general to personalised management, and the idea of prevention and treatment are examples of this, has changed the focus from disease treatment to preventive healthcare. The future development of modern medicine can be represented by these improvements, which put a strong emphasis on addressing patients' unique requirements while enhancing the effectiveness of medical care \cite{zakzouk2023blockchain}.

In the past decade, CV has received attention as an instrument for modern healthcare applications, paving the way towards advancements of visualization in the medical field \cite{jiang2023user}. CV in healthcare is an exhilarating and fast-developing field that comprises the use of computer algorithms mainly machine learning-based algorithms for the analysis of medical images and extract useful medical information. Medical images such as X-rays, CT scans, MRI, ultrasound, and others generate vast amounts of data that can be used to diagnose, monitor, and treat a wide range of diseases and conditions. Medical imaging encompasses a range of techniques used to produce images of the human body for therapeutic purposes, as well as for the detection, diagnosis, and analysis of diseases, and the study of normal anatomy and physiology \cite{ma2023segment}. The field of medical imaging has advanced from the initial use of X-rays to the more recent MRI technology, with a significant increase in the use of CV methods to support medical imaging progress \cite{parvaiz2023vision}. Examples of CV-based healthcare applications include detecting and diagnosing diseases such as cancer, identifying abnormalities through medical images, keeping track of the progress of treatment, monitoring vital signs, and more. As technology advances, it is likely that we will see even more innovative and impactful applications of CV in healthcare in the years to come.

Altameem and Ayman \cite{altameem2020facial} has proposed a facial recognition system to accurately determine the patient's facial expression in healthcare monitoring applications. The proposed system has three convolutional layer of CNN architecture to accurately recognize the face expression from the input captured from the patient's face. The first layer eliminates the faulty classification by ignoring the effects of textual and facial differentiators. The second convolutional layer validates the facial signs captured from the input against the output of the first layer. The last layer recognizes the mismatching error rates and stores in the training set. Thus, the proposed system used combination of ML algorithm, feature extraction and multimodal data fusion for facial visualization system. The experimental result shows 95.702\% accuracy when compared with conventional face recognition algorithms. Further, the applications of such CV techniques in healthcare applications leverage the robust and promising results. Integration of BC and CV in healthcare expand the opportunities like visual analysis \cite{wu2014thermal}, patient identity management \cite{leo2017computer}, clinical trials and research \cite{morinan2022computer}, and telemedicine \cite{li2022moving}.

CV has the potential to revolutionize healthcare, but there are limitations to its use in this field \cite{elyan2022computer}. One limitation is the limited availability and quality of data required to train and validate CV algorithms. Obtaining high-quality medical imaging data can be difficult and time-consuming, and the data may be biased or incomplete. Another limitation is the lack of interpretability of many CV algorithms, which can make it challenging for clinicians to trust the results and make informed decisions based on them. Additionally, medical conditions are often complex and multifaceted, making it challenging for algorithms to accurately diagnose and classify them. There are also ethical considerations, such as privacy and data security, and potential biases in the algorithms that could disproportionately affect certain populations. Finally, integrating CV into clinical workflows can be challenging and require collaboration between healthcare providers, data scientists, and technology experts \cite{gao2018computer}. Addressing these limitations is important to maximize the benefits of CV in healthcare.

BC technology can aid CV in healthcare by providing a secure and decentralised system for storing, sharing, and analysing medical images and data \cite{ramzan2022healthcare}. Here are some ways in which BC can help CV in healthcare:
Data Security: Medical images and data are sensitive and must be stored securely to prevent unauthorised access, hacking, or manipulation. BC technology uses cryptography and a distributed ledger system to ensure data privacy, security, and immutability \cite{esposito2018blockchain}.
Data Interoperability: In healthcare, there is a need for seamless interoperability of medical data and images across different platforms and institutions. BC can provide a platform for the standardisation and interoperability of medical data and images, which can improve collaboration and research across different healthcare providers \cite{gordon2018blockchain}.
Data Management: CV relies on large amounts of data to train algorithms and models. BC can help manage the storage, access, and sharing of data, allowing researchers and practitioners to access relevant data in a timely and efficient manner \cite{yaqoob2021blockchain}.
Data Traceability: Medical data and images often come from different sources, making it difficult to track the origin and authenticity of the data. BC can help track the provenance and authenticity of medical data and images, ensuring that they are reliable and trustworthy \cite{abbas2022blockchain}.
Data Sharing: Sharing medical data and images between healthcare providers can be challenging due to privacy and security concerns \cite{kumar2021integration}. BC can enable secure and transparent sharing of medical data and images, allowing healthcare providers to collaborate on diagnosis and treatment plans while maintaining patient privacy.

Thus, the BC-secured CV smart healthcare systems have the potential to revolutionize the healthcare industry by improving patient care and reducing costs. Further, BC-secured smart CV healthcare systems have the potential to improve patient outcomes and provide greater efficiency and security in the healthcare industry.

\subsection{Agriculture sector }

The conventional agricultural practices mainly focus on factors such as biodiversity, localization and shared genetic resources. Even though the traditional methods of agriculture have various advantages such as maximized food production, efficient land use, and easy adaptability, it has numerous downsides as well. Some of the major cons of the traditional methods include ruining the soil in the long run, spread of plant diseases, long-term pollination problems and so on. Smart farming plays a significant role in such scenarios by emphasizing more on the crop consistence, capital gain and production. Even though the agriculture sector has faced several challenges since the beginning, IoT-enabled smart farming brought noteworthy changes enabling precision agriculture, crop monitoring, livestock monitoring, irrigation and fertilizer management, soil quality analysis and smart pest control. Recent technological advancements in computer science have proved its significance in all aspects of human life including agriculture. Quality research has already been carried out by several researchers in almost all the major areas concerning smart agriculture. CV is one such advancement that made a huge positive impact on smart agriculture. Generally, CV is implemented as a 3-step process. The steps are 1. Image acquisition 2. Image processing and 3. Analyzing the image. Implementing CV technologies in agriculture has various applications ranging from saving production costs to enhancing productivity. It also helps in identifying the product defects, sort the produce based on various factors such as color, weight, size and so on. Some of the state-of-the-art approaches in the CV-enabled smart agriculture are discussed here. 

The authors in \cite{sunil2022weed} proposed a method for efficient classification of weed and crop species with the aid of CV under greenhouse conditions. Furthermore, to control the weeds in corn production, CV was adopted by \cite{sapkota2022uas}. CV can also be used for precision pollination by analyzing the movement and behaviour of insects. The authors in \cite{ratnayake2022spatial} has used techniques such as an automated and offline counting of insects, motion tracking and behavioral analysis. The phenology of specific crop can also be monitored with the street-level imagery using CV-based techniques \cite{d2022monitoring}. Similarly, the authors in \cite{sivaranjani2022overview} explored the use of CV in various post-production activities concerning agriculture such as grading and sorting. By analyzing the literature, it could be learned that even though adopting CV in agriculture has numerous benefits, there are various significant challenges as well. Both the food producers and consumers face many difficulties in agriculture and supply chain systems. The challenges include transparency between various partners, trust and connectivity between the various stakeholders, lack of confidence in food provenance and so on. 

BC can be regarded as an evolutionary next phase in the Information and Communication Technology (ICT) agriculture. The applications for smart agriculture using CV can further be improved using BC by storing and sharing the data, providing an audit trail, and enabling data verification. This will eventually enable peer-to-peer transactions to be carried out in a transparent manner, thus bypassing the need for any kind of intermediaries such as middleman in this sector. Instead of trusting a central authority in the agri-food market, the trust can be entirely on the cryptographic mechanisms and the peer-to-peer architecture. BC can thus be used for tracking every single information about the plants on various factors such as the quality of seed, the growth pattern of the crop and to understand the journey of any plant even after it leaves the farm. This will further help the concerned authorities to reward the producers who follow the best practices in agriculture. 

\subsection{Smart city development}

Conventional cities are complex because of organized people, businesses, transportation, communication networks, services, and utilities. The authors in \cite{ergazakis2004towards} conducted a detailed investigation of the problems faced by citizens in conventional cities and proposed a way towards knowledgeable cities. As a city grows, it generates technical, social, and organizational pressures that intimidate the economy and sustainability. For instance, no proper management of garbage and pollution in the cities. Other leading critical problems such as inadequate water management, inefficient electricity usage, poor traffic management, and high rates of crime. To address these challenges, conventional cities should transform and adopt technological developments such as CV, IoT, and distributed computing technologies to become smart cities. By taking on smart city technologies, conventional cities can enhance resource management, reduce waste and pollution, public safety, and promote economic growth \cite{iqbal2022blockchain}.

A smart city is a new urban development vision that brings together various sectors through the deployment of advanced technologies like IoT and distributed computing technologies. The primary goal is to enhance the city's existing resources by integrating, managing, and improving them. The focus is on enhancing processes in sectors such as healthcare, power, transportation, water management, education, commerce, and more. As the world's cities continue to grow larger and larger, younger generations emerge, and their preferences differ from older generations. Hyper-globalization is predominant worldwide, and the technological revolution is accelerating. Therefore, it is essential to create smart living environments that improve and enhance people’s quality of life, automate city services and processes, and develop transparent systems. Alternatively, the objective is to build smarter and more connected communities. The goal of any smart city should be to make sure that citizens' requirements are met and that they have the technological ways to access the city's services through a technology-driven and citizen-centered government approach.

AI has opened up new opportunities for creating sustainable cities, improving public services and monitoring urban infrastructure. This can be made possible by collecting information from various intelligent devices that are implemented for various purposes. One of the most significant services offered by AI to smart cities is security. In connection with this, CV technology is widely used to provide safety in smart cities, and it also has several applications such as critical infrastructure protection, sanitation and waste management, transport, traffic management, pandemic control, security, smart water management, and disaster management. These applications allow surveillance and decision-making processes through CV intelligence. Conversely, the security of the data generated by the CV systems remains an unsolved challenge. While CV systems, the data generated by these systems are still vulnerable to attack. So, an added security layer is needed to protect the information generated by the CV systems, which can be provided by BC technology.

\subsubsection{Critical infrastructure protection}
Smart cities make use of intelligent computing technologies to collect and analyze data from various sources to optimize resources, and monitor activities. These technologies also help prevent potential risks and improve services for the nation. Critical infrastructures are crucial resources for societal development, and failure or impairment of such infrastructure can have a significant impact on those who rely on it. This includes roads, water, communication, and energy systems. According to \cite{talari2017review}, security is the most crucial element of a smart city from the common man’s perspective. Any interference to the critical infrastructure, either accidental or deliberate, can lead to degraded system performance and result in social and economic losses.
Traditional video surveillance systems through Closed Circuit Television (CCTV) have become crucial for security and law implementation. However, they have their own limitations. Firstly, the person monitoring the displays can easily become frustrated by the numerous and simultaneous video streams in front. This may cause loss of focus on the incidents that happen \cite{surette2005thinking}. Additionally, multi-camera surveillance systems generate huge amounts of data, which demands high data rate and bandwidth for distributed processing. In addition to that, latency in communication is a significant problem for delay-sensitive applications. Finally, implementing an ad-hoc communication network with high data rates is expensive. It will be challenging to manage centrally after deployment \cite{cohen2009cctv}.

CCTV-based surveillance systems have restrictions for monitoring critical infrastructure due to the vast volume of data generated and the high latency communication delay. Therefore, fully automated real-time surveillance systems are essential to protect critical infrastructure in future smart cities. These surveillance systems must be adaptable to dynamic environments and bandwidth needs to react immediately when an event of interest occurs. The authors in \cite{isern2020reconfigurable} propose a secured re-configurable cyber-physical system that makes use of cloud, edge, and BC technologies for the protection of critical infrastructure \cite{gadekallu2021blockchain}. The local edge resource uses DL to recognize people, and high-performance system-on-chip (SoCs) embedded in the node process the data, achieving the real-time performance of 100 frames per second. This aids to manage the bandwidth demand of video streams from other camera sources at a lower frame rate. The cloud server gathers the information from the edge nodes to perform perimeter monitoring, tracking, and facial recognition. To make sure the integrity of the data during transmission and prevent manipulation, BC-based secured transmission is utilized. A quality and resource management unit monitors bandwidth and prompts reconfiguration to adapt to the transmitted video resolution. This system was tested in a real-time scenario, and the results showed a 75\% decline in bandwidth utilization compared to a no-reconfiguration scenario while data integrity is also maintained with the help of BC. Table \ref{bcsecuredvsstandalone} has given the comparison of traditional CV systems with BC secured CV systems based on the performance metrics considered.

\subsubsection{Traffic management}
It is widely acknowledged and accepted that the population is becoming more urbanized in recent years. It is predictable that by 2040, 70\% of the population will belong to urban areas. This migration of population will inevitably add more strains to the existing urban infrastructure. This includes flyovers, traffic signals, and subways, due to an increase in traffic flow. Appropriate traffic management can be crucial to avoid congestion and its related problems. An integrated intelligent traffic system is essential to locate and alleviate congestion areas. The gathered data can be used to create dynamic strategies that take preventive action before congestion occurs. Therefore, initiatives towards smart traffic management, are essential to improve the city's infrastructure. This includes dynamic traffic signals, finding appropriate parking slots, network-connected traffic signals, and other related components.

One of the most essential applications of smart city initiatives is finding a parking slot for vehicles. Double parking and repeated loading and unloading of goods have a huge impact on traffic management, so there is a need for real-time traffic data for the general public which may be helpful for various purposes like route decision, parking slot identification, and so on. In addition, an application to automatically find a parking slot based on current traffic conditions. The authors in \cite{Smartcity_traffic_01} proposed a CV-based roadside occupation surveillance system (CVROSS) to collect real-time traffic data through high-resolution cameras installed in cities. It captures images of loading and unloading activities happening on the roadside. Using the collected data, the system uses a visual representation of roadside occupancy and vacancy identified by fuzzy logic. This improves the transparency of roadside activities. However, the analyzed traffic data have to be shared with the public to help them make travel decisions. It is essential to protect the data before sharing it with the public. The authors in \cite{9437751} suggested using a data-sharing scheme based on BC. The proposed model combines the Ethereum BC with federated learning ideas and uses off-chain storage approaches to share data with the public. This model aids protect the information when shared with the public and ensures its integrity.

Smart traffic management system generates enormous amounts of data through installed cameras, enabling real-time recognition and monitoring of traffic incidents. Traditionally, the data collected is forwarded to the central server for in-depth analysis, which may lead to bottleneck and delay due to the demand for resources with computing abilities. To overcome this limitation, edge resources known as cloudlets have been proposed. The authors of \cite{Smartcity_traffic_06} focused on two traffic monitoring tasks, congestion monitoring, and speed detection, and designed a solution that consumes two algorithms, one implemented at the edge node and the other on the central server. However, the footage collected must be secured to prevent manipulation or adversarial attacks. BC technology has been proposed as a solution to secure footage and share it with relevant parties such as insurance organizations and law enforcement agencies. For instance, BC-based security frameworks \cite{Smartcity_traffic_05} can be used to store and share footage, ensuring its integrity and accessibility. Automatic damage assessment mechanisms based on video footage \cite{Smartcity_traffic_03} can also be used to prevent fraudulent insurance claims.

\subsubsection{Security}

Monitoring suspicious activity can be a challenging task with a wide range of applications, including video surveillance, anomaly detection, and intelligent transport systems. When a suspicious event happens recorded by a camera, searching through recordings to take action can be a delayed and time-consuming process. An automated method that identifies suspicious events in advance and alerts the appropriate authorities for immediate action can be more appealing and beneficial to society. While the traditional CCTV system can be used for event surveillance, it may not be possible to monitor CCTV events manually around the clock. In \cite{Smartcity_security_02}, the authors presented a hybrid model that uses the YOLO-V4 architecture to detect areas of interest where suspicious events are possible. The system sends information to the 3D-CNN architecture for activity recognition based on temporal information, allowing the detection of events by the surveillance camera and reporting to the authorities. However, those involved in such activities can be well aware of technological advances and the deployment of surveillance cameras. So, there can be a possibility to make adversarial attacks on the system before or after the event occurs. \cite{Smartcity_security_03} \cite{Smartcity_security_04} provide a detailed review of possible adversarial attacks that can collapse a CV system, such as perturbation attacks, black box attacks, and decision-based attacks. These attacks can make a well-trained model foolish and make it difficult to distinguish from the original image. To protect the system from such attacks, BC technology can be used. The event-based encryption proposed in \cite{Smartcity_security_01} aids prevent adversarial attacks on the system. The author suggests a decentralized data-sharing network powered by BC technology to ensure data undergo a vetting process before it is accepted by the network, providing a secure way to share sensitive information.

\subsection{Supply chain Monitoring, manufacturing and logistics }

Manufacturing is the process of creation or production of finished goods with the help of raw materials, human labor, tools, equipment, and machinery that is usually carried out systematically on a large scale. Supply chain management (SCM) is a process of managing the flow of goods, its associated data, and finances, starting from the raw materials all the way to delivering the product. Logistics is the part of SCM which takes care of mobilizing and storing the goods from the point of procurement to the point of delivery \cite{rehan2023supply}.

Systems like enterprise resource planning (ERP) and radio frequency identification (RFID) are being used in SCM, manufacturing, and logistics for several decades. ERP is a comprehensive system that deploys a common data model that benefits the business by enhancing transactions and resource planning. RFID are smart labels that provide ways to track products inside and outside the warehouse. However, these advancements still companies need human eyes in critical areas of SCM, manufacturing, and logistics. Unfortunately, employing a human requires consistent coordination of the eye and brain to perceive, interpret, act and respond back to the system. 
Interestingly, deploying a CV uses machine learning algorithms to identify, and classify objects based on the training model and discover patterns in images and videos. Self-driving car manufacturing companies like BMW, Tesla, and Volvo use CV to acquire image data from the environment so that the cars can detect objects, lane changes, signs, and traffic signals.  CV provides more flexibility in tracking the products than RFID, as during logistics there is no need to replace the broken or lost tags. Movement analysis enhances the logistics industry by streamlining the logistic processes through gesture recognition rather than entering the commands in the device manually. This would save time in acquiring the data and prevent errors from data entry operators. Using CV can help identify problems and optimize processes in SCM, manufacturing, and logistics.  

The authors, in this paper\cite{abosuliman2021computer}, presented effective solutions to logistics management by incorporating CV-based deep learning algorithms. The proposed methodology helps to determine the uncertainties in the delivery and in offering optimal logistical services.\cite{zhang2022logistics} The authors in this paper have proposed a scheme that will improve the traceability of the logistics through CV and image recognition techniques. The migration movements are monitored throughout the delivery points and warehouses by the proposed methodology that uses CV and image processing techniques.   \cite{liu2022computer} The authors in this paper have envisioned the use of CV in the process of bio-printing for bone research. Extrusion-based bio-printing is the current method that is widely used in the printing field. However, this method has a lesser printing resolution which is a remarkable limitation of  this method. CV in addition to the advanced development of high-resolution image processing, AI, and microelectronics addresses the problems of low resolution in extrusion-based bioprinting methods.

Advanced CV systems work by building AI models with images and videos. CV is considered as the highest form of AI and due to legal requirements, security features are increasingly important in CV-based supply chain management, manufacturing, and logistics. Edge AI vision systems allow private on-device processing in real-time, without sending visual data to the cloud. But for a robust system, collective data need to be shared for building deep learning models. For planning and predicting the performance of SCM and logistics, system simulation methods are important. Though the transactions in SCM and logistics are captured through modern image recognition algorithms, tampering with the original data is one of the inevitable limitations of CV.

BC is a digital ledger of transactions that is duplicated and distributed across the network. Essentially, decentralization is one of the BC characteristic features that would avoid in CV the need for a third party to be present to authenticate transactions and hence the concept of smartness and automation in devices are enabled. Utilizing BC in CV-based decision systems in SCM, manufacturing, and logistics would enhance robustness, as data is securely shared and a collective decision is obtained for an application. 
The image data collected through CV are used by the decision systems to train the model. Employing BC can increase the trust among the users. The immutability character of BC allows the data to be stored and processed in the third-party system also. Mutual trust among participants is one of the BC features that leverage the transaction verification process and delay in the delivery of the product in logistics. The traceability feature of BC helps to track the fleets in the logistics industry and the expiry status in the food and medicine supply chain.

\begin{table*}[t]
\centering
\begin{tabular}{|p{0.5cm}|p{1.2cm}|p{7cm}|p{7cm}|}
\hline
No & Performance Metric & BC secured CV Systems                                                                 & Standalone CV systems                                        \\ \hline
1.    & Speed              & Slower due to consensus and validation of transaction by all participants             & Can be optimized for real-time processing                    \\ \hline
2.    & Scalability        & BC scalability challenges may have an impact on processing of large scale image data. & Depends on architecture and resource of system               \\ \hline
3.    & Cost               & BC integration may involve additional cost for storage and computation recources.     & Depends on hardware, software and maintenance requirements.  \\ \hline
4.    & Storage            & BC storage may lead to redundant storage                                              & Can optimize storage based on specific requirements.         \\ \hline
5.    & Flexibility        & Smart contracts can be used to define rules and automated actions for captured image. & Depends on the capabilities and configuration of the system. \\ \hline
\end{tabular}
\caption{BC secured CV systems vs. standalone CV systems}
\label{bcsecuredvsstandalone}
\end{table*}

\section{Projects}
This section briefly introduces some well-known BC and CV integrated projects
\subsection{Everledger}
Everledger is an organisation that offers technological solutions to promote supply chain transparency globally. Their goal is to increase confidence and clarity in marketplaces where transparency is a fundamental requirement. The integration of BC and CV enables Everledger to offer an effective and transparent system for ensuring the authenticity and ownership of assets like diamonds throughout their lifecycle. The digital certificates stored on the BC are accessible and verifiable by a variety of stakeholders, including consumers, retailers, insurers, and law enforcement officers, thereby assisting in combating against fraud, tampering and stealing in the luxury goods market \cite{Everledger}.

\subsection{Provenance}
Provenance is a non-profit organisation that manages, supports, and finances an ecosystem for its community that facilitates the development and operation of complex BC-based DeFi applications. Using BC and CV, the Provenance platform provides a digital audit trace of a product's lifecycle, from its production to its consumption.  The technology uses CV to recognise and monitor individual wine bottles from the vineyard to the store. This helps prevent wine from being counterfeited or tampered with \cite{Provenance}. 

\subsection{Chainlink}
Chainlink is the Web3 services platform that connects the people, enterprises, and data of today to the Web3 world of tomorrow. it is a decentralised oracle network that provides smart contracts with data from the real world. The network collects data from a variety of sources, including weather stations, traffic cameras, and sensors, using CV. Smart contracts can use this information to make decisions, such as pricing insurance policies or triggering payments \cite{Chainlink}. 

\subsection{Ocean Protocol}
Ocean is an open-source protocol designed to facilitate the exchange and marketing of data and data-based services between organisations and individuals.Ocean Protocol employs BC technology to provide a secure and transparent data-sharing platform. Data providers can upload their data which will be encrypted and stored on the BC. The platform analyses data using CV and makes it more accessible to users. This information can be used for research, product development, and marketing, among other things \cite{Ocean}. 

\subsection{Grid+}
Grid+ is one of the BC-related energy businesses, has received \$29 million in pre-sales before ever going public. With the integration of BC and CV, the startup is eliminating the middleman in electricity transactions. This energy trading platform enables users to buy and sell energy directly from one another. The utilised BC technology ensures the immutability of transaction records, enhances security, and provides transparency for energy transactions. The platform monitors energy consumption and prices using CV. This information allows users to make informed choices decisions their energy consumption \cite{Grid+}. 
\section{Challenges and future directions}

\begin{figure*}[t]
	\centering
	\includegraphics[width=\linewidth]{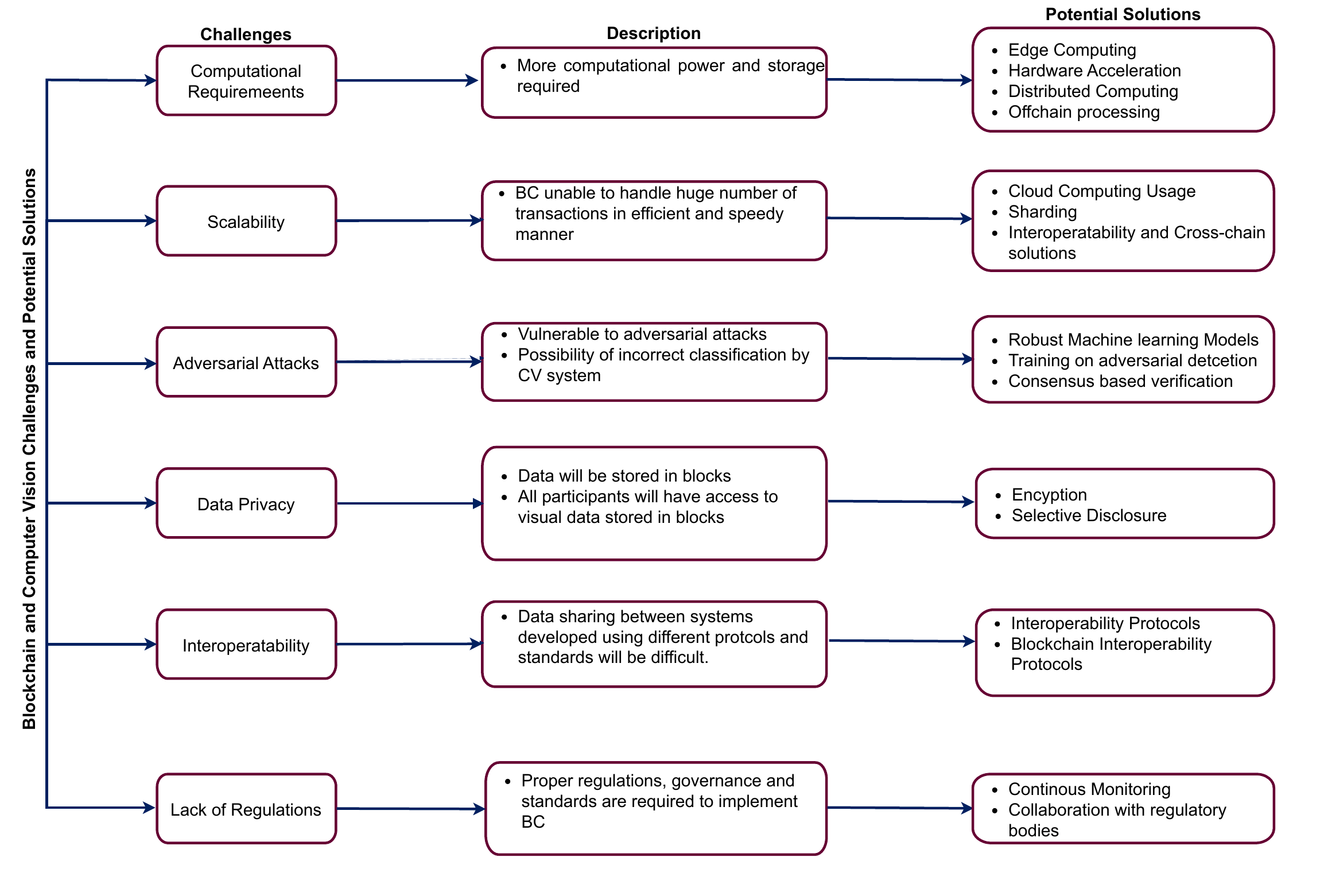}
	\caption{Challenges and Future Directions}
	\label{Challenges and Future Directions}
\end{figure*}

This section discusses about the challenges of integrating BC in CV and presents possible solutions to address the same. Fig.\ref{Challenges and Future Directions} has shown the identified challenges and future directions. 
\subsection{Computation requirements}
Integrating BC to the existing CV solutions may require significant changes in the hardware and software systems. Also, implementation costs would further increase due to the changes in the hardware and software, licensing, maintenance and so on, while integrating with the legacy systems. More storage and computational power would be required with the growing data size. Lack of specialist engineers is another significant challenge with respect to BC implementation. Some of the solutions for addressing such challenges involve a detailed need analysis on the use of BC in the system, and  providing specialized training to the professionals. Collaborative BCs with different working groups is yet another solution to deal with the existing implementation challenges. The idea is to understand the common challenges between industries and work in close collaboration to develop significant solutions that will benefit every party even without sharing any kind of proprietary information. Using BC as a service (BaaS) is another strategy that could be adopted by organizations with the help of which all the BC services can be used without having to invest more on the technical talent behind it.
\subsection{Lack of regulations and standards}
One of the significant challenges of BC-enabled CV solutions is the lack of interoperability between the wide range of BC networks. Lack of proper regulations and standards is yet another challenge that needs to be addressed while integrating BC to the CV-based systems. As there are no specific regulations, organizations need not follow any kind of rules while using BC in their business, that may result in a "state of disarray" due to which different BC networks fail to communicate with each other. Efficient consensus and better interoperability can be assured if there are a set of global standards and governance mechanisms. Failing to do so even compromises consistency from basic processes like security and making mass adoption. Some countries does not promote the use of BC technologies due to various security and environmental factors. However, with proper regulations, governance and standards, more organizations would start using BC in their CV systems. This not only will help the organizations to collaborate with each other on application development, but also in validating proofs of concept, easier integration with legacy systems and in sharing the various BC-based solutions.

\subsection{Scalability}
Scalability in BC can be regarded as its ability to handle huge number of transactions in an efficient and speedy manner. There is always a trade-off between scalability, security and decentralization, and can achieve only two out of the three among these simultaneously, and not all the three. This is referred to as "BC scalability trilemma". As CV encompasses image acquisition and analysis, BC scalability needs to be addressed. The significant issue occurs due to the fact that the BC requires all the participating entities to agree on the validity of the transactions. There are three basic approaches that could be adopted to solve the scalability issues in integrating BC with CV. The "on-chain" solutions require a change in the fundamental structure of the BC, whereas the "off-chain" solutions work by adding a second layer to the actual BC to speed up the transactions. Yet another solution involve the change in the consensus mechanism itself. FL is another advancement in technology innovations that significantly helps not only in preserving data privacy, but also in dealing with the scalability issues in the network. 

\subsection{Adversarial attacks}

Although BC technology has been adopted in CV, adversarial attacks remain a persistent threat. Adversarial attacks involve manipulating data or images to deceive CV system's, resulting in incorrect classifications. While BC can improve security and transparency, it may not fully address the CV system’s vulnerability to adversarial attacks. As a result, researchers and developers are exploring various techniques to enhance machine learning model resilience and robustness against adversarial attacks. Such techniques include developing models that are resistant to adversarial perturbations or using ensemble learning methods that combine multiple models to improve accuracy and reliability. In conclusion, while BC technology can assist in improving CV system security and transparency, it is not a comprehensive solution to address adversarial attack challenges. Additional measures and techniques are necessary to ensure the CV system’s resilience and robustness against such attacks.
\subsection{Data Privacy}
CV systems deal more with visual information that contains personal information about individuals. For instance, visual data on individuals includes images and videos during surveillance. Here, ensuring the privacy of captured visual data is crucial. BC integration requires careful attention to mechanisms to protect identities and prevent illegal access to information. Because, BC features of immutability and transparency make data stored in blocks and available to all participants, It is necessary to deploy robust encryption techniques to securely store visual data on the BC, to ensure authorised individuals are accessing the information.

\subsection{Interoperatability}
Interoperability is a critical challenge when it comes to adopting BC technology in CV. This challenge arises because different BC networks may use different protocols and standards, making it difficult for different systems to share and interpret data accurately and efficiently. This lack of interoperability can limit the scalability and usefulness of the technology. For instance, suppose a CV system is developed on one BC platform that uses specific standards and protocols. In that case, it may not be compatible with another BC platform that uses different standards and protocols. Consequently, exchanging data and models between the two systems can be challenging, reducing the potential applications of the technology. To tackle this challenge, researchers and developers are exploring ways to establish common standards and protocols for BC adoption in CV. This can involve developing open-source frameworks and protocols that can be used across different BC networks, allowing for easier data exchange and interoperability. By promoting interoperability between different BC platforms, it may be possible to overcome some of the challenges associated with the adoption of BC technology in CV.
\section{Conclusion}

In conclusion, the integration of BC with CV has the huge potential to transform the sectors such as healthcare, defense, agriculture, smart city and supply chain management. This potential can be utilized to enable the development of innovation applications. By integrating the decentralized and secure nature of BC with visual analytics ability of CV, new opportunities emerge to achieve better transparency, traceability, and accountability.  Though, this integration has proven results in sectors like supply chain, healthcare, and surveillance systems, challenges and open issues such as scalability, data bias, adversarial attacks, computational complexity, data privacy and interoperability still unaddressed. With continued efforts, advancements and new research initiatives can reshape this combination which can help to drive transformative solutions that benefit society as a whole.

\bibliographystyle{IEEEtran}
\bibliography{main.bib}
\end{document}